# Reinterpreting the Origin of Bifurcation and Chaos by Urbanization Dynamics

Yanguang Chen

(Department of Geography, College of Urban and Environmental Sciences, Peking University, Beijing 100871, PR China. Email: chenyg@pku.edu.cn)

**Abstract:** Chaos associated with bifurcation makes a new science, but the origin and essence of chaos are not yet clear. Based on the well-known logistic map, chaos used to be regarded as intrinsic randomicity of determinate dynamics systems. However, urbanization dynamics indicates new explanation about it. Using mathematical derivation, numerical computation, and empirical analysis, we can explore chaotic dynamics of urbanization. The key is the formula of urbanization level. The urbanization curve can be described with the logistic function, which can be transformed into 1-dimensional map and thus produce bifurcation and chaos. On the other hand, the logistic model of urbanization curve can be derived from the rural-urban population interaction model, and the rural-urban interaction model can be discretized to a 2-dimensional map. An interesting finding is that the 2-dimensional rural-urban coupling map can create the same bifurcation and chaos patterns as those from the 1-dimensional logistic map. This suggests that the urban bifurcation and chaos come from spatial interaction between rural and urban population rather than pure intrinsic randomicity of determinate models. This discovery provides a new way of looking at origin and essence of bifurcation and chaos. By analogy with urbanization models, the classical predator-prey interaction model can be developed to interpret the complex dynamics of the logistic map in physical and social sciences.

# 1. Introduction

Chaos is one of the important subjects of Science in the twentieth Century. However, the problems of origin and essence of chaos were not really solved in last century, and they are passed on to the new century. The simplest model for understanding chaos is the well-known logistic map. The complicated behavior of the logistic growth brought to light by May (1976) led to a profound insight into complex dynamics. Thus, chaos is always regarded as intrinsic randomicity of determinate dynamical systems. However, why simple mathematical models can exhibit very complicated behavior? Or what dynamical mechanism can be revealed from behind the logistic equation? This is still a pending question. Fortunately, a two-population interaction model of urbanization helps us answer this question now and gives us a further insight into bifurcation and chaos. The work of May (1976) is based on the 1-dimension logistic map. A numerical simulation will show in this work that the same period-doubling bifurcation and chaos can be produced by using a 2-dimensional map of rural-urban interaction. In urban geography, the urbanization curve is always characterized by the logistic equation, which is associated with the rural-urban population migration model. The period-doubling cascade and chaotic behavior can be derived from the two-population interaction process of urban systems, and the result is identical in images to the patterns coming from the one-population logistic equation based on ecosystems. This model offers us with a new angle of view to understand complex behaviors of simple dynamical systems.

Urbanization provides new way of understanding the origin and essence of chaos. Urban systems are complex systems, and the process of urbanization and urban evolution are nonlinear process associated with chaos and fractals (Albeverio *et al*, 2008; Batty, 1995; Chen, 2008; Dendrinos and El Naschie, 1994; Portugali, 2011; Wilson, 2000). Using mathematical derivation, numerical computation, and empirical analysis, we can reveal new knowledge about bifurcation and chaos based on the nonlinear dynamics of urban evolution. A basic and important measurement of urbanization is the proportion of urban population to the total population, which is termed "level of urbanization" in urban geography. The curve of urbanization level can be described with sigmoid functions such as logistic function, which can be discretized to a 1-dimensional map. Using the formula of urbanization level, we can derive the logistic function from the rural-urban population interaction model, which can be discretized to a 2-dimensional map. Thus the 1-dimensional logistic map can be associated with the 2-dimensional rural-urban interaction map. As will be shown below, the 2-dimensional rural-urban map can create the bifurcation and chaos that are identical in patterns to those produced by the 1-dimensional logistic map. This suggests that the origin of bifurcation and chaos is two-population coupling and interaction rather than intrinsic randomicity of determinate models (Chen, 2009).

The study of chaos associated with bifurcation can help us understand natural and social systems deeply. This paper is a development based on a series of previous studies (Chen, 2009; Chen, 2012; Chen, 2014). The rest of this work is organized as follows. In Section 2, the bifurcation and chaos from rural-urban population interaction dynamics are illustrated by using a 2-dimensional map, and a phase portrait analysis of rural-urban interaction is performed. In Section 3, an empirical analysis is made by means of American census data to verify the rural-urban interaction model. The case study lays the foundation of experiments for the urbanization model. In Section 4, several related questions are discussed. First, the two-population interaction model is generalized to explain the ecological phenomena including logistic growth and oscillations of population. Second, the scaling laws of period-doubling cascade are compared with those of hierarchy of cities. Third, the nonlinear dynamics of urbanization curve is further generalized to the fractal dimension curve of urban growth. Fourth, the nonlinear replacement dynamics is outlined. Finally, the discussion is concluded with a brief summary.

## 2. Mathematical models

### 2.1 The two-population interaction model

A rural-urban population interaction model can lead to a new understanding of chaos. This model is made by means of observations and statistical data (Chen, 2008). Under certain condition of postulates, the rural-urban model for two interacting population group is given as follows (Chen, 2009a)

$$\begin{cases} \dfrac{dr(t)}{dt} = ar(t) - b\dfrac{r(t)u(t)}{r(t)+u(t)} \\ \dfrac{du(t)}{dt} = c\dfrac{r(t)u(t)}{r(t)+u(t)} \end{cases}, \qquad (1)$$

where $r(t)$ and $u(t)$ refers to rural population and urban population at time $t$, respectively ($r(t)>0$,

$u(t)>0$), *a*, *b*, and *c* are parameters. The model shows that the spatial interaction between urban and rural population causes urbanization, namely, transition of rural population into urban population in a region. According to equation (1), the growth rate of rural population depends on rural population size and the two-population interaction, while that of urban population only depends on the rural-urban interaction. If the study region is a close system, then the parameters *b* and *c* are equal to one another, i.e. $b=c$, or else they are not. Equation (1) has a firm basis of statistical analysis. The model can be verified with the population data set of American census since 1790.

It can be proved that the system of differential equations on rural-urban interaction is equivalent to the logistic equation of urbanization curve. Rewriting equation (1) in a simple form, we have a general model of urbanization dynamics such as

$$\begin{cases} \dfrac{\mathrm{d}r(t)}{\mathrm{d}t} = r(t)[a - b^* u(t)] \\ \dfrac{\mathrm{d}u(t)}{\mathrm{d}t} = c^* r(t) u(t) \end{cases}, \quad (2)$$

where

$$b^*(t) = \frac{b}{r(t)+u(t)}, \quad c^*(t) = \frac{c}{r(t)+u(t)}. \quad (3)$$

The level of urbanization can be expressed as

$$L(t) = \frac{u(t)}{P(t)} = \frac{u(t)}{r(t)+u(t)} = 1 - \frac{r(t)}{r(t)+u(t)}, \quad (4)$$

in which $L(t)$ denotes urbanization level at time *t* (obviously $0 \leq L(t) \leq 1$). The level of urbanization is an important measurement in urban study. Just because of the definition of urbanization level, the 1-dimension map of logistic growth can be associated with the 2-dimension map of rural-urban interaction. In fact, taking the derivative of the equation (4) yields

$$\frac{\mathrm{d}L(t)}{\mathrm{d}t} = \frac{\mathrm{d}u(t)/\mathrm{d}t}{r(t)+u(t)} - \frac{u(t)}{[r(t)+u(t)]^2}\left[\frac{\mathrm{d}r(t)}{\mathrm{d}t} + \frac{\mathrm{d}u(t)}{\mathrm{d}t}\right]. \quad (5)$$

Substituting equation (2) into equation (5) gives

$$\frac{\mathrm{d}L(t)}{\mathrm{d}t} = \frac{c^* r(t) u(t)}{r(t)+u(t)} - \frac{u(t)}{[r(t)+u(t)]^2}\left[ar(t) - (b^* - c^*)r(t)u(t)\right]. \quad (6)$$

For simplicity, we can postulate that the region is a close system, which has no population exchanged with outside. In this case, we have $b=c$, $b^*=c^*$, and thus we have

$$\frac{dL(t)}{dt} = \frac{c^* r(t)u(t)}{r(t)+u(t)} - \frac{ar(t)u(t)}{[r(t)+u(t)]^2} = c^* r(t)L(t)\left[1 - \frac{a}{c^* u(t)}L(t)\right]. \tag{7}$$

As indicated above, $c^*=c/[r(t)+u(t)]$, equation (7) can be reduced to

$$\frac{dL(t)}{dt} = c\frac{r(t)}{r(t)+u(t)}L(t)\left[1 - \frac{a}{cu(t)/[r(t)+u(t)]}L(t)\right]. \tag{8}$$

Based on the level of urbanization defined by equation (4), the logistic equation is readily derived as below

$$\frac{dL(t)}{dt} = c(1-\frac{a}{c})L(t)\left[1 - \frac{u(t)}{r(t)+u(t)}\right] = (c-a)L(t)\left[1 - L(t)\right]. \tag{9}$$

In literature, equation (9) is always expressed as follows

$$\frac{dL(t)}{dt} = kL(t)\left[1 - L(t)\right], \tag{10}$$

where $k=b-a=c-a$ represents the intrinsic/original rate of growth. The discrete expression of equation (10) is identical in form to the simple model employed by May (1976) to research complicated dynamics. By using the 1-dimension map based on equation (10), we can produce period-doubling bifurcation and chaotic patterns through numerical simulation, which have been explored by many scientists who are interested in chaos and complexity.

## 2.2 Bifurcation and chaos based on 2-dimensional map

Discretizing the rural-urban population interaction model yields a 2-dimensional maps, which can be employed to make numerical analysis. Since equation (10) can be derived from equation (1) through mathematical transformations, we expect that the complicated dynamical behaviors such as period-doubling oscillation and chaos can also be produced by using the maps based on equation (1). First of all, equation (1) should be discretized and converted into the following form

$$\begin{cases} r(t+1) = (1+\alpha)r(t) - \beta\dfrac{r(t)u(t)}{r(t)+u(t)} \\ u(t+1) = u(t) + \gamma\dfrac{r(t)u(t)}{r(t)+u(t)} \end{cases}, \tag{11}$$

in which the parameters $\alpha$, $\beta$, and $\gamma$ for discrete form correspond to $a$, $b$, and $c$ for continuous form in equation (1), respectively. The parameters in equation (11) will vary slightly after continuous-discrete transformation. If the region is a close system, we will have $\beta=\gamma$. The first parameter can be taken as *α*=0.025 according to the US urbanization model. In addition, the initial values of rural population and urban population can be taken as *r*(0)=3.727559 million and *u*(0)=0.201655 million in terms of the American census in 1790. Thus, the dynamics of spatial interaction and transition between rural and urban population can be numerically simulated by means of computer. As expected, the behaviors of the 2-dimension maps defined by equation (11) are indeed identical in feature to the complicated dynamics of the 1-dimension logistic map in ecology (Figure 1).

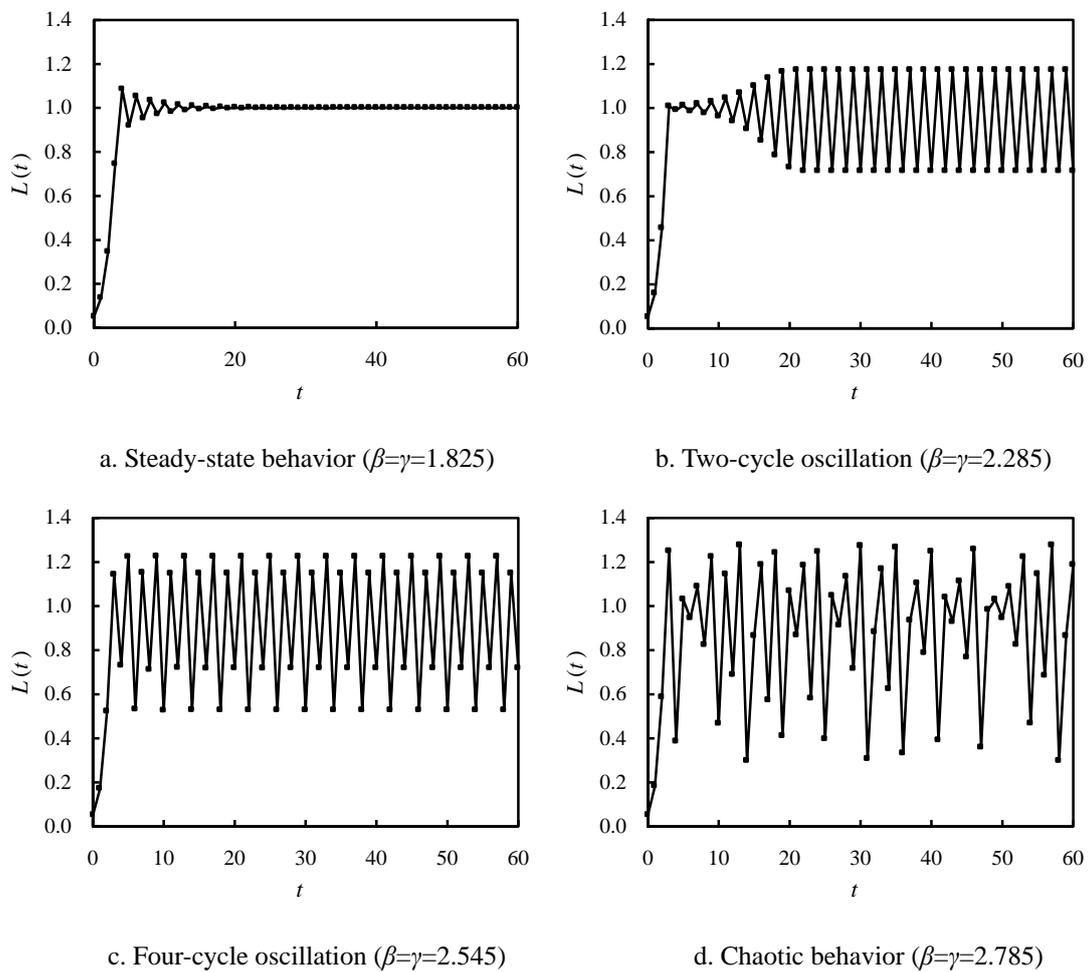

a. Steady-state behavior (*β*=*γ*=1.825)　　　　b. Two-cycle oscillation (*β*=*γ*=2.285)

c. Four-cycle oscillation (*β*=*γ*=2.545)　　　　d. Chaotic behavior (*β*=*γ*=2.785)

**Figure1 The growth process of level of urbanization simulated by the 2-dimension maps of rural-urban interaction model: from steady state to chaos**

(**Note**: The parameter values of the model would be given as *a*=0.025, *r*(0)=3.727559 million and *u*(0)=0.201655 million. In addition, the limitation, *β*=*γ*, is necessary for corresponding to the 1-dimension logistic map. See Chen (2009))

The main results can be summarized as follows. If *β*=*γ*<0.025, the urbanization level displays a stable equilibrium decay to zero ($L_{max}$=0), while if 0.025<$β$=$γ$<1.032, it shows a stable equilibrium growth to unit ($L_{min}$=1). The latter implies the standard logistic growth curve which represents a type of trend lines of urbanization level in the real world. If 1.033<$β$=$γ$<2.025, the urbanization level shows stable alternating change to unit (Figure 1a); when 2.025<$β$=$γ$<2.475, the system will fall into a stable cycle of period 2 (Figure 1b); when 2.475<$β$=$γ$<2.571, a stable cycle of period 4 appears (Figure 1c); and then, with *b* and *c* increasing, a stable cycle of period 8 (*b*=*c*>2.571), and 16 (*b*=*c*>2.591), and even $2^n$ (here *n*>4 is a positive integer) is exhibited. Lastly, when $β$=$γ$>2.61, the system comes into chaotic state. The limit of the parameters is about $β$=$γ$=3.033 (Chen, 2009a)

An interesting finding comes from the comparison between 2-dimensional maps based on rural-urban interaction and the 1-dimensional map based on logistic growth. In fact, equation (10) can be discretized as $L(t+1)=(1+K)L(t)-KL(t)^2$, where the parameter *K* corresponds to the parameter *k* in equation (10). Then we have $K \propto β-α$. Comparing the above results with those of May (1976) shows the common characters of behaviors shared by the 2-dimension map of rural-urban population interaction model and the 1-dimension map of the logistic growth model. Moreover, the critical parameter values of period-doubling bifurcation and chaos of the interaction model are almost equivalent to those of the logistic model but a subtle difference. Numerical simulation experiments demonstrate that when *α* becomes very small, the critical value of the period-doubling route to chaos proceeding from the rural-urban interaction will be approximately equal to that directly from the logistic equation. This discovery indicates a new way of looking at the origin and essence of bifurcation and chaos.

Another finding is the intrinsic relation between order and chaos. There are narrow ranges of periodic solutions in the chaotic "band". When $β$=$γ$>2.857, the level of urbanization evolves into a cycle of period 3, further when $β$=$γ$>2.871, appears a cycle of period 5, and then period 6 or even period 7, in short, a non-$2^n$-cycle. However, if $β$=$γ$>2.88, the level of urbanization will gradually evolve into a non-periodic state again. The periodic behaviors in chaos differ from those outside the chaotic belt. The former is of non-$2^n$-cycle, while the latter of $2^n$-cycle. All the non-$2^n$-cycles, including the period 5 or the period 6, suggest a randomness indicating chaos just as period 3 does (Li and Yorke, 1975). In this sense, this sort of limited chaos seems to be defined as "random + non-doubling period" (Chen, 2009a). One the other hand, some slight disorder can be found during the $2^n$-cycles, which can be revealed by spectral analysis. This suggests that chaos and order cannot be absolutely separated, and they contain one another or are interweaved each other.

It is easy to examine the sensitive dependence on the initial conditions of the level of urbanization when the system evolves into the chaotic state. For example, let's take $\alpha=0.025$ and $\beta=\gamma=2.785$. Then change the American rural population in 1790 from $r(0)=3.727559$ million to $r(0)=3.727558$ million, that is, reduce one rural person, and the urban population is unchanged. The numerical experimental results are displayed in Figure 2, which contains two curves of urbanization level: one is under the unchanged initial conditions mentioned above, and the other, under the changed initial conditions. Clearly, the difference between the two curves becomes larger and larger with the increasing iteration times.

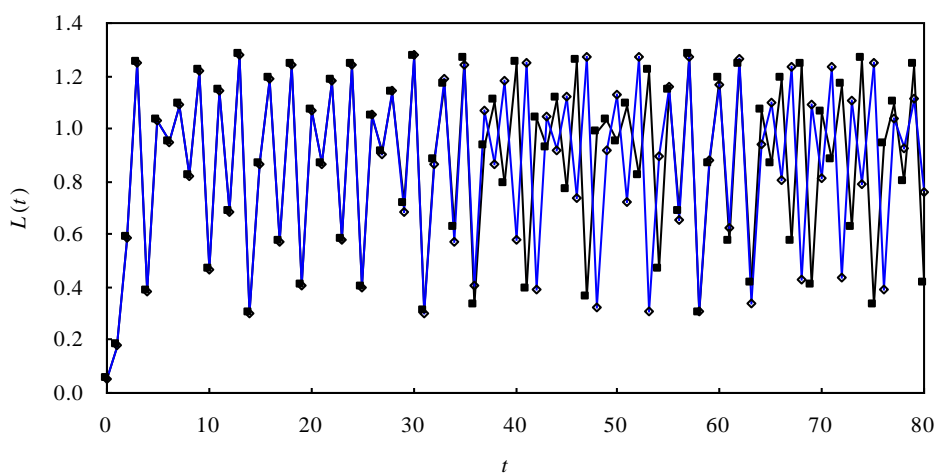

**Figure 2 Sensitive dependence on initial conditions of the urbanization in chaotic state**
(**Note**: The square solid dots refer to the estimated values under the original initial conditions, whereas the rhombic hollow dots to the estimated values under the changed initial conditions.)

However, where urbanization is concerned, bifurcation and chaos mostly occur in the possible world rather than the real world. The parameter value $\alpha=0.025$, which is based on the American urbanization data, is of practical sense. In this case, the level of urbanization only make sense if the other two parameters satisfy the condition such as $0.025<\beta=\gamma<1.032$. In other words, the urbanization level will exceed 1 and both the rural and urban population will present negative values if $\beta=\gamma>1.032$. It is these parameters with no physical meaning for urbanization that lead to period-doubling bifurcation and chaos. Whether or not urbanization in the real world can exhibit bifurcation and chaos is still a pending question.

## 2.3 Phase portraits of 2-dimension map

Using the 2-dimensional map, we can draw the phase portraits of the logistic process based on the 1-dimensional map. The spatio-temporal feature of urbanization dynamics can be revealed with the phase portraits. If we take rural population $r(t)$ as $x$-axis and urban population $u(t)$ as $y$-axis, we can make the scatterplots of period-doubling bifurcation and chaotic behavior of urbanization process in phase space with equation (11). Thus the period-doubling can be reflected by $2^n$ ($n=1,2,3,…$) radials whose crosspoint is the original point in this plot ($r(t)=0$, $u(t)=0$)(Figure 3a-c). If the urbanization level evolves into a chaotic state, all points distribute randomly in the region confined by the two intersectant radials (Figure 3d). Sometimes there are several or even numerous radials indicating non-$2^n$-cycle in the plot. No essential difference appears between the phase portrait based on 2000 iterations and that based on 20000 iterations. However, those points don't converge. In other words, no strange attractor emerges in the chaotic pattern.

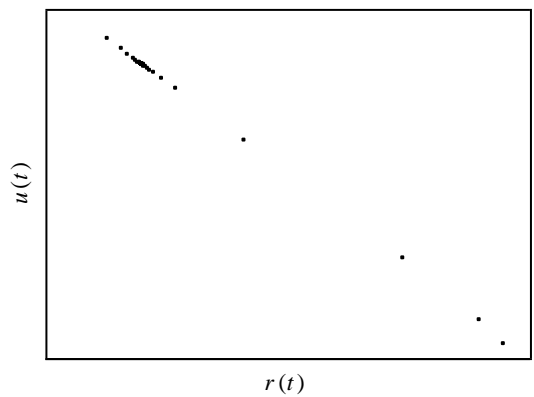
a. Steady state ($β=γ=1.825$)

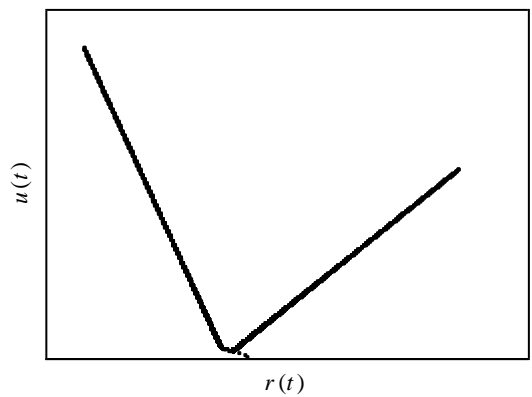
b. Stable cycle of period 2 ($β=γ=2.285$)

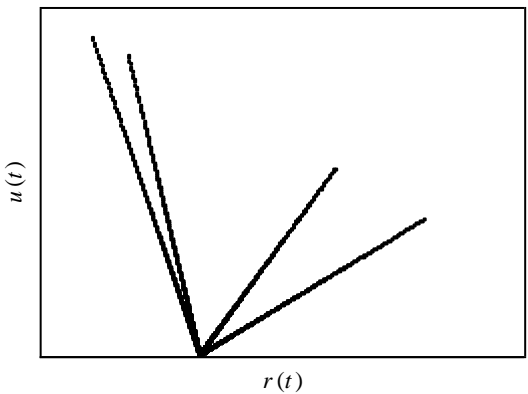
c. Stable cycle of period 4 ($β=γ=2.545$)

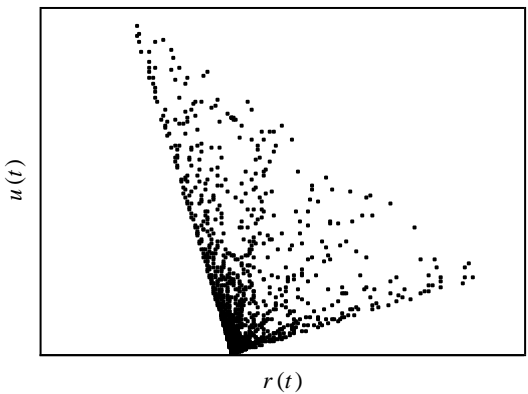
d. Chaotic state ($β=γ=2.785$)

**Figure 3 The phase portrait of the period-doubling bifurcation and chaos of rural-urban population interaction**

(**Note**: The times of iterations is 2500. The four subplots in Figure 3 correspond to the four subplots in Figure 1, respectively. See Chen (2009))

Despite the fact that no chaotic attractor can be found, these scatter points follow certain mathematical rule. The distance from a data point ($r(t)$, $u(t)$) to the origin (0, 0), i.e., the crosspoint of radicals which act as boundaries of these points, can be defined as

$$d = [r(t)^2 + u(t)^2]^{1/2}. \tag{12}$$

Then the distribution of these points satisfy the following logarithmic relation

$$N(d) = A \ln d - B. \tag{13}$$

where $N(d)$ denotes the accumulative number of points within distance $d$, and $A$ and $B$ are parameters. Now, let's investigate mathematical structure of the phase space from the angle of view of statistics. Suppose the distance is $d=4^n$, where $n$ is a natural number, and the number of iterations is 5000. For example, if $\beta=\gamma=2.785$ as given, the estimated value of parameters are $A=369.87$ and $B=509.33$, the regression degree of freedom is $df=8$, and the goodness of fit is $R^2=0.9999$ (Figure 4). Changing the parameter values of equation (11) result in different values of A and B, but the logarithmic relation will not vary. For instance, if $\beta=\gamma=3$, then the parameters of equation (13) will be $A=156.11$ and $B=228.93$, accordingly $df=22$, $R^2=0.9997$.

The derivative of the logarithmic function is a hyperbolic function. This implies that the density of the points in the phase portrait of chaotic state declines outwards from the origin in the form of a hyperbolic curve. Although both cities as systems and systems of cities are fractals (Batty, 1995; Batty and Longley, 1994; Chen, 2008; Frankhauser, 1994), the phase portrait of the chaos coming from rural-urban population interaction shows no self-similar pattern. The inverse function of the logarithmic function is the exponential function. We can learn the property of the logarithmic model from the exponential model. Compared with the normal distribution, the exponential distribution suggests complexity (Goldenfeld and Kadanoff, 1999); but compared with exponential distribution, the power-law distribution indicates complexity (Barabási, 2002; Barabási and Bonabeau, 2003). This implies that complexity is a relative conception. Exponential distribution comes between simplicity and complexity. By analogy with the exponential function, the logarithmic distribution of the points in the phase portrait of chaos suggests a pattern coming between simple structure and complex structure.

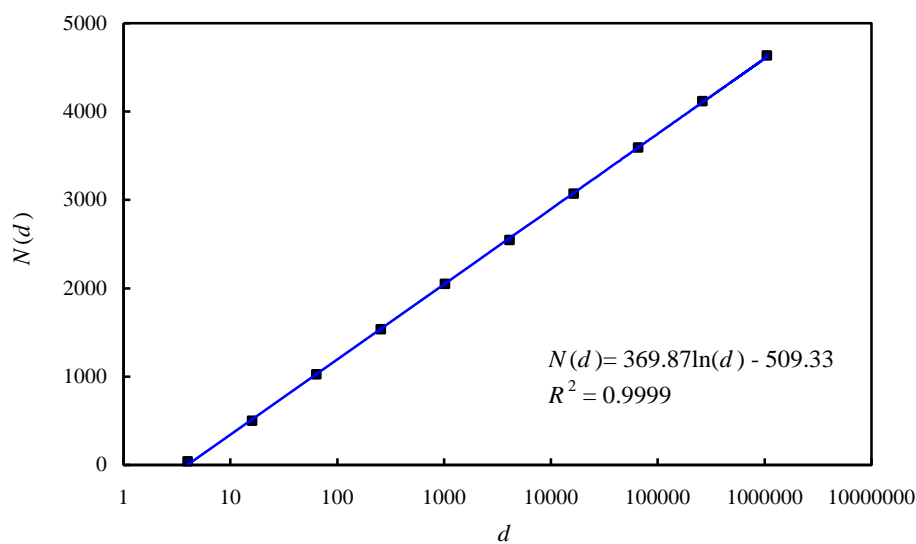

**Figure 4 The structural character in the phase space of the chaos of urbanization**
(**Note**: The plot corresponds to the fourth subplot in Figure 3, and the parameter values are $\beta=\gamma=2.785$.)

## 3. Empirical analysis

### 3.1 Data and method

The numerical experiments are based on the 2-dimensional map from the rural-urban population interaction model. It is necessary to make empirical analysis using the dynamical equations of urbanization. There are two central variables in the study of spatial dynamics of city development: population and wealth (Dendrinos, 1992). According to aim of this study, only the first variable, population, is chosen to test the models. In fact, population represents the first dynamics of urban evolution (Arbesman, 2012). Generally speaking, the population measure falls roughly into four categories: rural population $r(t)$, urban population $u(t)$, total population $P(t)= r(t)+u(t)$, and the level of urbanization indicative of the ratio of urban population to the total population, $L(t)= u(t)/P(t)$.

The American data comes from the population censuses with the interval of about 10 years. Although the website of American population census offers 23 times of census data from 1790 to 2010, I only use the data from 1790 to 1960 (Table 1). The reason is that the US changed the definition of cities in 1950, and the new definition came into effect in 1970. From 1970 on, the American urban population was measured by the new standard. As a result, the statistic caliber of the population data from 1970 to 2000 might be different from those before 1970 although they approximately join with each other. The data displayed in Table 1 are fitted to the discretization expressions of the United Nations (1980) model and the Lotka-Volterra-type model respectively (r.e. Dendrinos and Mullally, 1985; Lotka, 1956; Volterra, 1931). The parameters of models are made by the least squares calculation, which can make the key parameters, slopes, fall into the most reasonable range. After estimating the model parameters, we should make tests in two ways. One is the well-known statistical tests, and the other is the logical tests, which is often ignored in literature. If the model fails to pass the statistical tests, it has problems such as incomplete or redundant variables, or inaccurate parameter values; if the model cannot pass the logical tests, it has structure problem so that it cannot explain the phenomena at present and predict the developing trend in future. Statistical tests can be made in definite procedure, while the logical tests needs to be done with the help of mathematical transform and numerical analyses.

Table 1 The US rural and urban population and the related data (1790-1960)

| Time (year) [$t$] | Interval (years) [$\Delta t$] | Rural population [$r(t)$] | Urban population [$u(t)$] | $\dfrac{r(t)u(t)}{r(t)+u(t)}$ | Rural rate of growth [$\Delta r(t)$] | Urban rate of growth [$\Delta u(t)$] |
|---|---|---|---|---|---|---|
| 1790 | 10 | 3727559 | 201655 | 191305.67 | 125855.30 | 12071.60 |
| 1800 | 10 | 4986112 | 322371 | 302794.21 | 172831.00 | 20308.80 |
| 1810 | 10 | 6714422 | 525459 | 487322.03 | 223077.60 | 16779.60 |
| 1820 | 9.8125 | 8945198 | 693255 | 643391.97 | 284153.58 | 44228.48 |
| 1830 | 10 | 11733455 | 1127247 | 1028443.23 | 348484.30 | 71780.80 |
| 1840 | 10 | 15218298 | 1845055 | 1645549.78 | 439908.20 | 172944.10 |
| 1850 | 10 | 19617380 | 3574496 | 3023569.39 | 560942.30 | 264202.20 |
| 1860 | 10 | 25226803 | 6216518 | 4987478.10 | 342920.70 | 368584.30 |
| 1870 | 10 | 28656010 | 9902361 | 7359287.97 | 740346.40 | 422737.40 |
| 1880 | 10 | 36059474 | 14129735 | 10151800.00 | 481402.70 | 797653.00 |
| 1890 | 10 | 40873501 | 22106265 | 14346837.12 | 512383.50 | 810856.70 |
| 1900 | 9.7917 | 45997336 | 30214832 | 18235956.49 | 425582.20 | 1210127.90 |
| 1910 | 9.7917 | 50164495 | 42064001 | 22879255.97 | 163788.26 | 1244862.74 |
| 1920 | 10.25 | 51768255 | 54253282 | 26490822.68 | 221831.22 | 1454372.39 |
| 1930 | 10 | 54042025 | 69160599 | 30336844.29 | 341720.60 | 554473.90 |
| 1940 | 10 | 57459231 | 74705338 | 32478532.68 | 373837.30 | 1542285.60 |
| 1950 | 10 | 61197604 | 90128194 | 36448706.03 | 506197.80 | 2293539.90 |
| 1960 | 10 | 66259582 | 113063593 | 41776788.81 | | |

**Source**: http://www.census.gov/population.

## 3.2 Parameters estimation and model selection

The above stated model on rural-urban interaction is an equation system coming from empirical analysis. One of the general forms of urbanization dynamics models can be expressed as

$$\begin{cases} \dfrac{dr(t)}{dt} = ar(t) + \varphi u(t) - b\dfrac{r(t)u(t)}{r(t)+u(t)} \\ \dfrac{du(t)}{dt} = \omega r(t) + \psi u(t) + c\dfrac{r(t)u(t)}{r(t)+u(t)} \end{cases} \quad (14)$$

This is in fact the urbanization model of United Nations (1980), in which $a$, $b$, $c$, $\varphi$, $\psi$, $\omega$ are parameters. In order to make statistical analysis based on the observational data, we must discretize equation (14) so that it transform from differential equations into difference expressions, i.e., a 2-dimension map. Then the analysis of continuous dynamics changes to that of discrete dynamics. If the time difference $\Delta t=10$ as taken, then $dx/dt \propto \Delta x/\Delta t$. Let $r(t)$, $u(t)$ and $r(t)*u(t)/[r(t)+u(t)]$ be independent variables, and $\Delta u(t)/\Delta t$ and $\Delta r(t)/\Delta t$ be dependent variables. The model can be fitted to the American census data of rural and urban population. A multivariate stepwise regression analysis based on the least squares calculation gives the following model

$$\begin{cases} \dfrac{\Delta r(t)}{\Delta t} = 0.02584 r(t) - 0.03615\dfrac{r(t)u(t)}{r(t)+u(t)} \\ \dfrac{\Delta u(t)}{\Delta t} = 0.05044\dfrac{r(t)u(t)}{r(t)+u(t)} \end{cases} \quad (15)$$

This is a pair of difference equations of which all kinds of statistics including $F$ statistic, $P$ value (or $t$ statistic), variance inflation factor (VIF) value and Durbin-Watson (DW) value can pass the tests at the significance level of $\alpha=0.01$. In this model, $\varphi=\psi=\omega=0$. Although we should have $b=c$ in theory, they are not equal in the empirical results. Two factors account for this. One is that the US is not a truly closed system because of mass foreign migration; the other is that the natural growth of the urban population depends on the rural-urban interaction. The second reason might be more important. But on the whole, the equations as a special case of the United Nations model can better describe the American rural and urban population migration process in the recent 200 years.

To examine the relationship between the 1-dimensional map and the 2-dimensional map of urbanization, we can investigate the US urbanization curve. According to equation (9), the level of urbanization should follow the logistic curve. It is easy to calculate the urbanization ratio using the data in Table 1. For convenience, we set time dummy $t$=year-1790. A least squares computation involving the percentage urban data gives the following results

$$L(t) = \frac{1}{1 + 20.41573 e^{-0.02238 t}}. \tag{16}$$

The goodness of fit is about $R^2$=0.9839. Thus we have $k$=0.02238 as the estimated value of the intrinsic growth rate. On the other hand, we could estimate the original rate of growth $k$ value by equation (15): one is $k_1$=$b$-$a$≈0.03615-0.02584=0.01031, and the other is $k_2$=$d$-$a$≈0.05044-0.02584 = 0.02460. The intrinsic growth rate should come into between $k_1$=0.01031 and $k_2$=0.02460 and indeed it does. The parameter values estimated from the dynamical system model, equation (15), are similar to that from the logistic model, equation (16). There are some differences between different estimated results due mainly to three factors. The first is non-closed region, the second rests with imprecise data, and the third lies in the computation error resulting from transformation from continuous equation to discrete expression.

For comparison and selection, we also fit the American rural and urban data to the discretization of the predator-prey interaction model. Let $r(t)$, $u(t)$ and $r(t)*u(t)$ be independent variables and $\Delta u(t)/\Delta t$ or $\Delta r(t)/\Delta t$ dependent variables. The multivariable stepwise regression based on least squares computation gives an abnormal result, which cannot be accepted (Chen and Xu, 2010). If we lower the requirements, then the American urbanization process could be expressed with the Keyfitz-Rogers model (Keyfitz, 1980; Rogers, 1968). However, this mathematical expression has two vital shortcomings, which defies us to accept the Keyfitz-Rogers model for the US urbanization (Chen and Xu, 2010). In short, neither the linear Keyfitz-Rogers model nor the usual non-linear Lotka-Volterra model is as good as the United Nations model in terms of logic sense and statistic effect.

### 3.3 Numerical experiment

As a complement analysis, the data of American urban, rural and total population and the urbanization level can be generated by using discrete dynamics model. And then, we can draw a comparison between the simulation value and observed data. Figures 5 and 6 show the simulation results based on equations (15), respectively. It can be seen that the change of the urban and total population approximately follow the path of the S-shaped curve, while the rural population first increases, then decreases, and finally turns itself into the urban population completely (Figure 5).

Moreover, the level of urbanization increases in the logistic way (Figure 6). The change trend of the numerical simulation results displayed in Figures 5 and 6 is roughly coincident with the actual observation data (Chen, 2008; Chen and Xu, 2010). Although it is unpractical that the capacity value of the urbanization level is 100%, the evolvement characters of the rural and urban population reflected by the discrete dynamical model, i.e., equation (15), comply with the logic rules of urbanization. The total population converges, and the change of the percentage urban takes on a logistic curve.

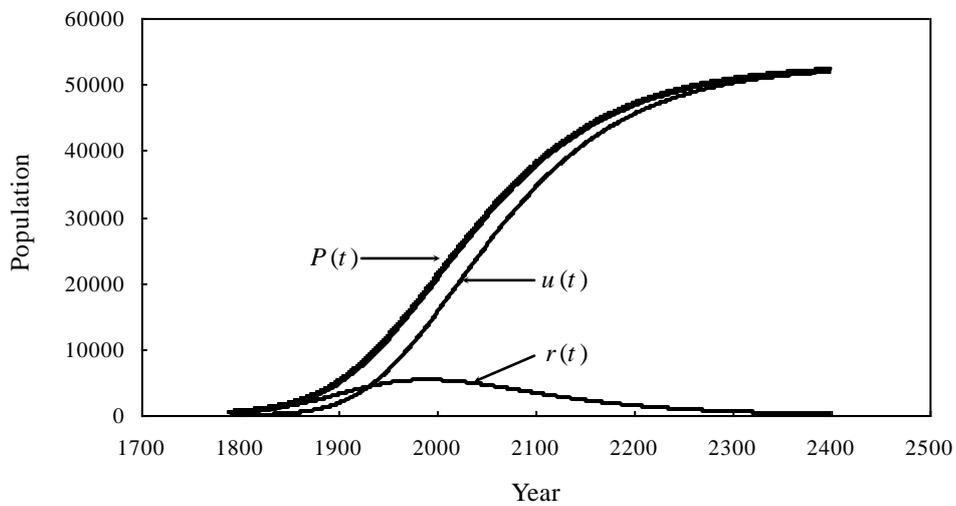

**Figure 5 The numerical simulation curve of rural, urban, and total population in the American urbanization process**

(**Notes**: The numerical simulation results are based on the discrete dynamical equations of urbanization, equation (15), the unit of population is taken as 10,000 persons. See Chen and Xu (2010))

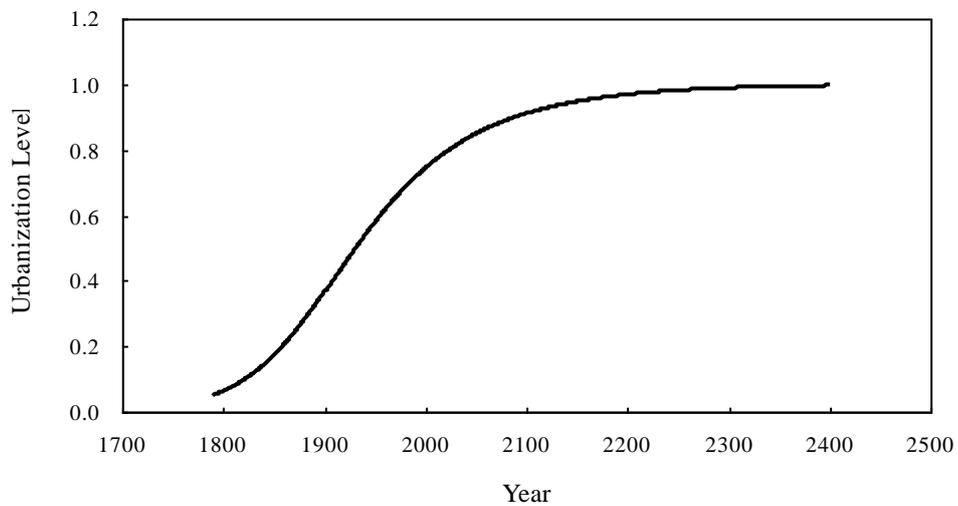

**Figure 6 The numerical simulation curve of American urbanization level (1790-2400)**

(**Notes**: The numerical simulation based on equation (15). The capacity value is 1. The curve is identical in shape to that of logistic growth indicated by equation (16). See Chen and Xu (2010))

So far, we have finished the building work of the model of urbanization based on the population observation in the real world. To sum up, the calculation results lend empirical support to the theoretical models and relations. First, the rural-urban population interaction model is testified, at least for a number of developed countries. The American model of rural-urban population interaction can be expressed by equation (1). This is the experimental foundation of theoretical analysis of discrete urbanization dynamics. Second, the relationship between the 1-dimensional map of logistic growth and the 2-dimensional map of rural-urban interaction is verified. By using the system of rural-urban interaction models, we can produce the logistic curve of urbanization. What is more, the curves of urban population, rural population, and total population are empirically acceptable. In the following section, I will discuss the related questions about bifurcation, chaos, complexity, and scaling law from the theoretical angle of view.

## 4. Questions and discussion

### 4.1 Generalization and supposition

According to the theoretical derivation, numerical experiments, and empirical analysis, chaos originates from nonlinear interaction between two coupling elements. The reasons are as below. First, a 1-dimensional logistic map is actually based on a 2-dimensional interaction map between two populations. Second, both the 1-dimensional map and the 2-dimensional map processes can create the same patterns of bifurcation and chaos. Further, the theoretical findings can be generalized to the other scientific fields. Urban systems are analogous to ecosystems in dynamical behaviors (Dendrinos, 1992). Besides the logistic equation, the predator-prey interaction model can be also applied to urban dynamics (Dendrinos and Mullally, 1985). The predator-prey system has several forms and can produce bifurcation and chaos (Chen, 2009b; Jing and Wang, 2006; Krise and Choudhury, 2003; Wang and Zeng, 2007). The period-doubling bifurcation and chaos coming from the 2-dimension map of urbanization dynamics remind us of the patterns of complicated dynamics from the 1-dimension logistic map of insect population. Moreover, the rural-urban interaction reminds us of the predator-prey interaction and the Lotka-Volterra model (Volterra, 1931; Lotka, 1956). By analogy, the conclusions drawn from urban models can be generalized to ecological field and *vice versa* (Table 2). A conjecture is that the logistic growth in ecology results from the two-population interaction. The well-known Lotka-Volterra model of ecological interaction can be revised as follows (Chen, 2009a)

$$\begin{cases} \dfrac{dx(t)}{dt} = ax(t) - b\dfrac{x(t)y(t)}{x(t)+y(t)} \\ \dfrac{dy(t)}{dt} = c\dfrac{x(t)y(t)}{x(t)+y(t)} - dy(t) \end{cases}, \qquad (17)$$

where $x(t)$ refers to the number of preys at time $t$, and $y(t)$ to the number of predators. The notation $a$, $b$, $c$ and $d$ are all constants. Equation (17) is in fact a generalized predator-prey interaction model. If $x(t)+y(t)=constant$ as given, then equation (17) will return to the original Lotka-Volterra model. The dynamical behaviors of equation (17) are similar to those of equation (1). Obviously equation (1) is a special case of equation (17). Defining the percentage of predator population as

$$z(t) = \dfrac{y(t)}{x(t)+y(t)}, \qquad (18)$$

we can derive a logistic equation from equation (17) such as

$$\dfrac{dz(t)}{dt} = (c-a-d)z(t)[1-z(t)]. \qquad (19)$$

From equation (19), a 1-dimension map of logistic growth can be obtained in the form

$$z(t) = (k+1)z(t-1) - kz(t-1)^2, \qquad (20)$$

in which the parameter $k \propto c-a-d$. Both the 1-dimension map stemming from equation (19) and the 2-dimension map coming from equation (17) can exhibit the same complicated dynamics as shown in Figure (1). This suggests that the two-population interaction accounts for the logistic growth, oscillations, and chaotic behavior in ecosystems.

**Table 2 The typical dynamical equations modeling urban evolvement and ecological phenomena**

| Model | Dynamical equation | Urban system | Ecosystem |
| --- | --- | --- | --- |
| Allometric growth | $\begin{cases} dx(t)/dt = ax(t) \\ dy(t)/dt = by(t) \end{cases}$ | Allometric scaling relations | Two-population competition |

| Two-population interaction | $\begin{cases} dx(t)/dt = ax(t) - bx(t)y(t) \\ dy(t)/dt = cx(t)y(t) - dy(t) \end{cases}$ | The rural-urban interaction | The predator-prey interaction |
|---|---|---|---|
| Generalized two-population interaction | $\begin{cases} \dfrac{dx(t)}{dt} = ax(t) - b\dfrac{x(t)y(t)}{x(t)+y(t)} \\ \dfrac{dy(t)}{dt} = c\dfrac{x(t)y(t)}{x(t)+y(t)} - dy(t) \end{cases}$ | The rural-urban interaction and logistic growth | Two-population competition and predator-prey interaction |

**Notes**: (1) In the equations, *a*, *b*, *c* and *d* are parameters. (2) The equations of allometric growth indicate simplicity, while the two-population interaction models suggest complexity.

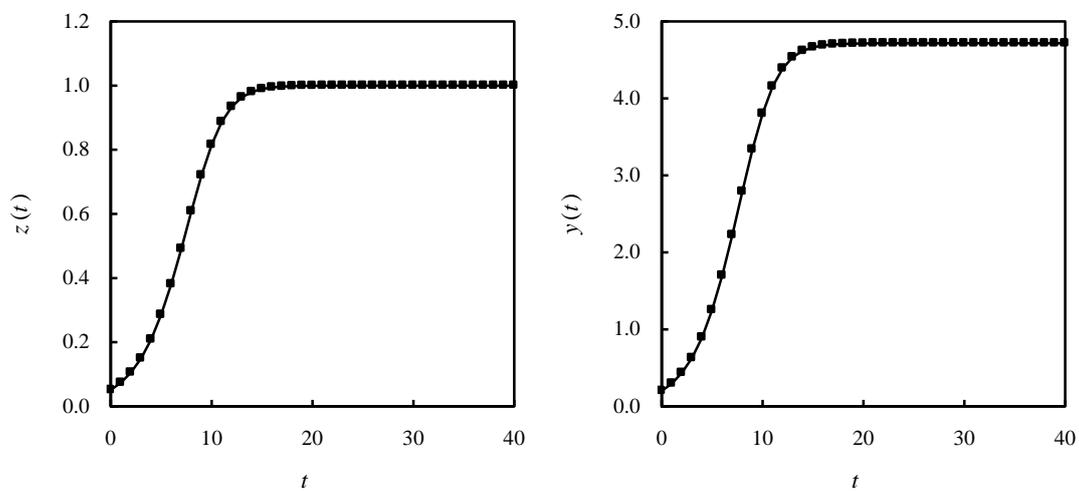

a. Logistic growth of percentage population      b. Quasi-logistic growth of predator population

**Figure 7 The logistic growth of percentage of predator population and quasi-logistic growth of predator population**

(**Note**: If the percentage of predator population follows the logistic growth mathematically (a), the predator population growth will exhibit a sigmoid curve seemingly (b). See Chen (2009))

A new supposition is that the logistic growth of population is only an approximate description. It is the percentage of predator population defined by equation (18), $z(t)$, rather than the predator population itself, $y(t)$, that follows the law of logistic growth. A numerical simulation based on the 2-dimension map originating from equation (17) shows that if the percentage of predator population $z(t)$ satisfies a sigmoid curve, the predator population $y(t)$ displays a sigmoid curve of growth in appearance (Figure 7). However, the latter is a fractional or quadratic logistic growth rather than the common logistic growth. Moreover, the oscillations of population, $x(t)$, $y(t)$, or even the total population $P(t)=x(t)+y(t)$, may reflect the period-doubling bifurcation and chaos of percentage population, $z(t)$. In short, the generalized predator-prey interaction can interpret more ecological phenomena than the classical Lotka-Volterra model (Chen, 2009a). Anyway, the studies on urban chaos can help us understand the John Holland's question. After discussing the Lotka-Volterra model, Holland (1995) said: "In the long run, extensions of such models should help us understand why predator-prey interaction exhibit strong oscillations, whereas the interactions that form a city are typically more stable."

## 4.2 Bifurcation diagrams and scaling laws

It is revealing to compare the period-doubling bifurcation process of urbanization dynamics with hierarchical structure of cities mathematically. Numerical simulations show that Feigenbaum's number, which is a universal constant brought out by Feigenbaum (1978), can be figured out by means of the rural-urban interaction model. Bifurcation diagram and tent map and such can be obtained, too (Figure 8). In fact, if we give away the assumption of regional close system, we will have $\beta \neq \gamma$ in equation (11). Thus we can find more complex dynamical behaviors exhibited by the two-population interaction model.

The cascade structure of urban systems and the period-doubling bifurcation route to chaos share the same scaling law. The period-doubling bifurcation diagram can be described with a set of exponential laws as follows

$$N_m = N_1 r^{m-1}, \tag{21}$$

$$L_m = L_1 \delta^{1-m}, \tag{22}$$

$$W_m = W_1 a^{1-m}, \tag{23}$$

where $m$ denotes the order of hierarchy of bifurcation, $N_m$ refers to period number (or bifurcation number), $L_m$ is the range for the stable periodicity, and $W_m$ is the span between two bifurcation points of order $m$. As for the parameters, $N_1=1$, $L_1$ and $W_1$ are constants, $r=2$, $\delta \approx 4.6692$, $a \approx 2.5029$ (Feigenbaum, 1978; Feigenbaum, 1979; Feigenbaum, 1980). In fact, equation (22) represents "the bifurcation-rate scaling law", and equation (23), "the fork-width scaling law" (Willliams, 1997). Accordingly, equation (21) represents "period-number scaling law". These what is called scaling laws are linear scaling laws, but they can be transformed into nonlinear scaling laws, i.e., power laws (Chen, 2008). An allometric scaling relation between the periodical range ($L_m$) and the fork width ($W_m$) can be derived from equations (22) and (23) and the result is

$$L_m = \mu W_m^b, \qquad (24)$$

in which the proportionality coefficient $\mu = L_1 W_1^{-b}$, and the scaling exponent can be defined by

$$b = \frac{\ln \delta}{\ln a} \approx 1.6796. \qquad (25)$$

The physical meaning of this number is not yet clear for the time being.

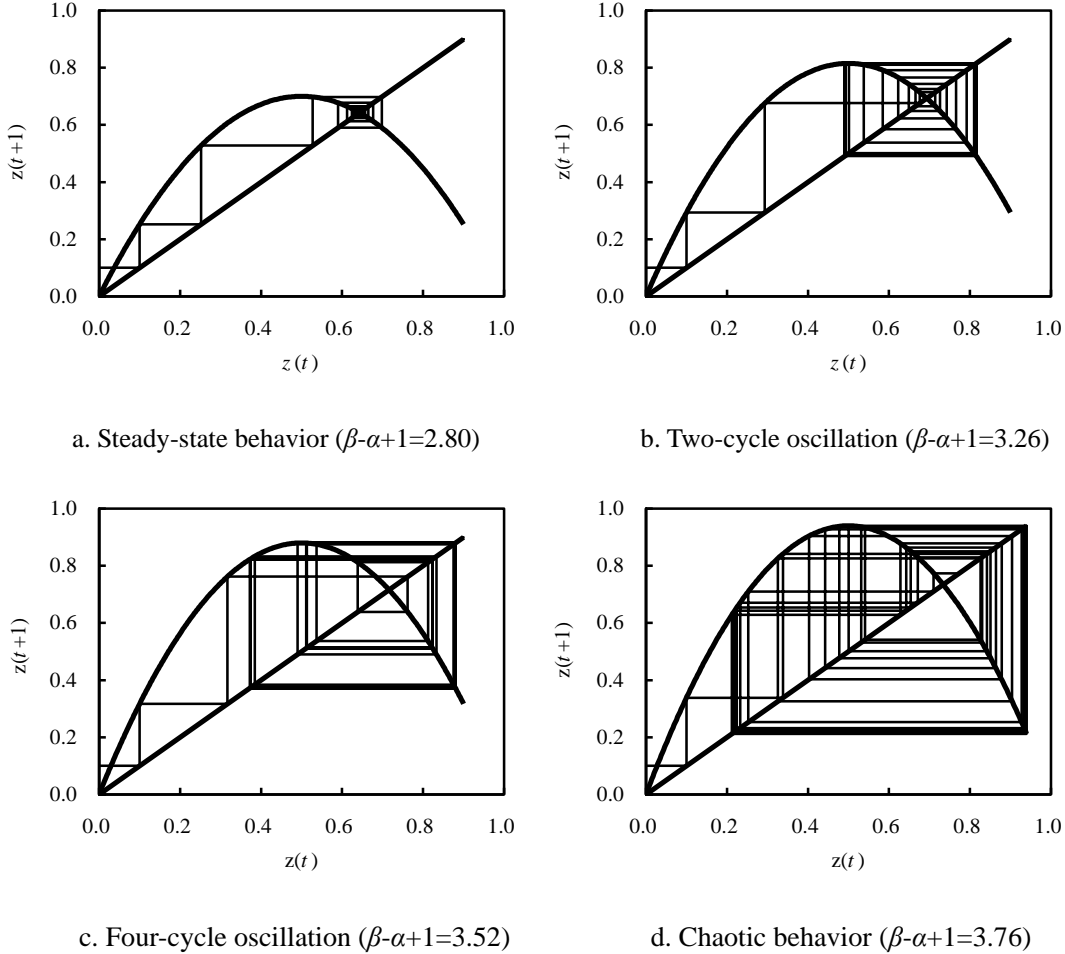

a. Steady-state behavior ($\beta-\alpha+1=2.80$)   b. Two-cycle oscillation ($\beta-\alpha+1=3.26$)

c. Four-cycle oscillation ($\beta-\alpha+1=3.52$)   d. Chaotic behavior ($\beta-\alpha+1=3.76$)

**Figure 8 Tent map: from steady state to chaos (the initial value is $L_0=0.01$)**

(**Note**: The graph of tent map is also termed "spider diagram", which can be seen in literature such as Messel (1985). The diagrams are created by using the 2-dimensional rural-urban interaction map based on equation (11). These subplots correspond to the subplots in Figures 1 and 3.)

The three exponential equations reflect the universal cascade structure of nature and society. There exists an analogy between the scaling laws of the period-doubling bifurcation and those of hierarchy of cities (Table 3). The period-doubling bifurcation scaling laws can be used to characterize the urbanization process, while the similar scaling laws can be employed to describe the cascade structure of urban hierarchies (Chen2011; Chen, 2012a; Chen, 2012b; Chen, 2012c; Jiang and Yao, 2010). What is more, the scaling relation, equation (24), is analogous with the allometric relationships between urban area and population. The law of allometric growth originally indicates that the rate of relative growth of an organ is a constant fraction of the rate of relative growth of the total organism (Beckmann, 1958; Bertalanffy, 1968; Lee, 1989). In urban geography, it can be used to describe the scaling relation between the urbanized area ($A_m$) of a city and its population ($P_m$) in the urban area (Batty and Longley, 1994; Chen, 2008; Lee, 1989). This analogy and comparability lend further support to the suggestion that the evolvement of urban systems comes between order and chaos (White and Engelen, 1994).

Table 3 Comparison between the linear scaling laws of period-doubling bifurcation and the exponential laws of hierarchy of cities

| Linear scaling law | Period-doubling bifurcation | Hierarchy of cities |
|---|---|---|
| The first law--number law | $N_m = N_1 r^{m-1}$ | $N_m = N_1 r_n^{m-1}$ |
| The second law—length/size law | $L_m = L_1 \delta^{1-m}$ | $P_m = P_1 r_p^{1-m}$ |
| The third law—width/area law | $W_m = W_1 a^{1-m}$ | $A_m = A_1 r_a^{1-m}$ |

**Notes**: (1) The scaling laws of hierarchy of cities are illuminated by Chen (2008). (2) The period-doubling bifurcation in this work comes from the 2-dimension map based on the rural-urban interaction model, which differs from the 1-dimension logistic map in form.

## 4.3 Dynamics of fractal dimension evolution of urban growth

The nonlinear dynamics of urbanization corresponds to the complex dynamics of urban growth and morphology. Urban growth can be measured with the time series of fractal dimension of urban form. The common fractal dimension can be obtained by box-counting method. In theory, the box dimension of urban form ranges from 0 to 2. However, in practice, the box dimension always comes between 1 and 2. Boltzmann's equation can be employed to describe the fractal dimension growth of cities (Chen, 2012d). In fact, Boltzmann's equation was used to model urban population evolution by Benguigui et al (2001). Urban population is associated with urban form and urbanization. The Boltzmann model of fractal dimension evolution is as follows

$$D(t) = D_{\min} + \frac{D_{\max} - D_{\min}}{1 + [\frac{D_{\max} - D_{(0)}}{D_{(0)} - D_{\min}}]e^{-kt}} = D_{\min} + \frac{D_{\max} - D_{\min}}{1 + \exp(-\frac{t - t_0}{p})}, \quad (26)$$

where $D(t)$ refers to the fractal dimension of urban form in time of $t$, $D_{(0)}$ to the fractal dimension in the initial time/year, $D_{\max} \leq 2$ to upper limit of fractal dimension, i.e. the capacity of fractal dimension, $D_{\min} \geq 0$ to the lower limit of fractal dimension, $p$ is a scaling parameter associated with the initial growth rate $k$, and $t_0$, a temporal translational parameter indicative of a critical time, when the rate of fractal dimension growth indicating city growth reaches its peak. The scale and scaling parameters can be respectively defined by $p=1/k$, $t_0=\ln[(D_{\max}-D_{(0)})/(D_{(0)}-D_{\min})]^p$. For the normalized variable of fractal dimension, equation (26) can be re-expressed as a logistic function

$$D^*(t) = \frac{D(t) - D_{\min}}{D_{\max} - D_{\min}} = \frac{1}{1 + (1/D_{(0)}^* - 1)e^{-kt}}, \quad (27)$$

where $D_{(0)}^* = (D_{(0)} - D_{\min})/(D_{\max} - D_{\min})$ denotes the normalized result of $D_{(0)}$, the original value of fractal dimension. Empirically, equations (26) and (27) can be supported and thus validated by the dataset of London from Batty and Longley (1994), the datasets of Tel Aviv from Benguigui *et al* (2000), and the dataset of Baltimore from Shen (2002). The derivative of equation (27) is just the logistic equation

$$\frac{dD^*(t)}{dt} = kD^*(t)[1 - D^*(t)]. \quad (28)$$

Without loss of generality, let the time interval $\Delta t=1$. Thus, discretizing equation (28) yields a 1-dimensional map such as

$$D_{t+1}^* = (1+k)D_t^* - kD_t^{*2}. \quad (29)$$

Defining $D_t^* = (1+k)x_t/k$, we can transform equation (29) into the following form

$$x_{t+1} = (1+k)x_t(1-x_t) = \mu x_t(1-x_t). \quad (30)$$

where $x_t$ is the substitute of $D_t^*$, and $\mu=k+1$ is a growth rate parameter. Equation (30) is just the well-known logistic map (May, 1976). If the fractal dimension of urban form can be fitted to Boltzmann's equation, it implies that urban evolution can be associated with spatial chaotic dynamics.

The process of urban growth is a dynamic process of urban space filling. An urban region falls into two parts: filled space and unfilled space. We can define a spatial *filled-unfilled ratio* (FUR) for urban growth, that is

$$O = \frac{D^*}{1-D^*} = \frac{U}{V}. \tag{31}$$

Thus we have

$$D^* = \frac{O}{O+1} = \frac{U}{U+V} = \frac{U}{S}, \tag{32}$$

where $U$ refers to the filled space area with various buildings (space-filling area), measured by the pixel number of built-up land on digital maps, and $V$, to the unfilled space area without any construction or artificial structures (space-saving area). Thus the total space of urbanized region is $S=U+V$. Obviously, the higher the $O$ value is, the higher the degree of urban spatial filling will be. The normalized fractal dimension can be termed *level of space filling* (SFL) of cities, implying the degree of spatial replacement.

Based on a digital map with given resolution, the filled space can be measured with the pixels indicating urban and rural built-up area such as structures, outbuildings, and service areas. In contrast, the unfilled space is the complement of the filled space of built-up area. On the digital map, the unfilled space is just the blank space of an urban figure. If a region is extensively developed and is already occupied by various urban infrastructure and superstructure, it is transformed, and the unfilled space is replaced by filled space. This spatial replacement dynamics can be described by a pair of differential equations

$$\begin{cases} \dfrac{dU(t)}{dt} = \alpha U(t) + \beta \dfrac{U(t)V(t)}{U(t)+V(t)} \\ \dfrac{dV(t)}{dt} = \lambda V(t) - \beta \dfrac{U(t)V(t)}{U(t)+V(t)} \end{cases}, \tag{33}$$

where $\alpha$, $\beta$, and $\lambda$ are parameters. This implies that the growth rate of filled space, $dU(t)/dt$, is proportional to the size of filled space, $U(t)$, and the coupling between filled and unfilled space, but not directly related to unfilled space size; the growth rate of unfilled space, $dV(t)/dt$, is proportional to the size of unfilled space, $V(t)$, and the coupling between unfilled and filled space, but not directly related to filled space size. From equation (33), we can derive equation (28). Discretizing equation (28) yields a 2-dimensional map of urban growth, which can be used to created periodic oscillation chaos similar to the patterns shown in Figure 1 (Chen, 2012c).

## 4.4 Replacement dynamics

The logistic growth model and the rural-urban interaction model can be employed to develop the theory replacement dynamics. Dynamical replacement is one of the ubiquitous general empirical observations across the individual sciences, which cannot be understood in the set of references developed within the certain scientific domain. We can find the replacement processes associated with competition everywhere in nature and society. The theory of replacement dynamics should be developed in the interdisciplinary perspective. It deals with the replacement of one activity by another. One typical substitution is the replacement of old technology by new, another typical substitution is the replacement of rural population by urban population. Urbanization is a process of population replacement, that is, the urban population substitutes for the rural population (Karmeshu, 1988; Rao *et al*, 1989). The components in a self-organized system, generally speaking, can be distributed into two classes, and the process of a system's evolution is a process of discarding one kind of component in favor of another kind of component. This process is termed "replacement" (Chen, 2012d; Chen, 2014). For example, the population in a geographical region can be divided into urban population and rural population, and urbanization is a process of rural-urban replacement of population (Rao *et al*, 1989); the technologies can be divided into new ones and old ones, and technical innovation is a process of new-old technology replacement (Fisher and Pry, 1971; Hermann and Montroll, 1972). In fact, people can be divided into the rich and the poor, the geographical space can be divided into natural space and human space, and so on. Where there are self-organized systems, there is evolution; and where there is evolution, there is replacement. Replacement results from competition and results in evolution. Replacement analysis is a good approach to understanding complex systems and complexity.

The basic and simplest mathematical model of replacement is the logistic function, which can be employed to describe the processes of growth and conversion. Besides, other sigmoid functions such as the quadratic logistic function and Boltzmann's equation may be adopted to model the replacement dynamics. A number of mathematical methods such as allometric scaling can be applied to analyzing various types of replacement. In fact, the allometric scaling can be used to analyze the relationships between the one thing/group (e.g. urban population) and another thing/group (e.g. rural population). A replacement process is always associated with the nonlinear dynamics described by two-group interaction model. The discrete expression of the nonlinear differential equation of replacement is a 1-dimensional map, which is equivalent to a 2-dimensional map. The maps can generate various simple and complex behaviors including S-shaped growth, periodic oscillations, and chaos. If the rate of replacement is lower, the growth curve is a sigmoid curve. However, if the replacement rate is too high, periodic oscillations or even chaos will arise. This suggests, no matter what kind of replacement it is--virtuous substitution or vicious substitution, the rate of replacement should be befittingly controlled. Otherwise, catastrophic events may take place, and the system will likely fall apart. The studies on the replacement dynamics are revealing for us to understand the evolution in nature and society, and the relationship between the 1-dimension map and the 2-dimension map is revealing for our understanding of the replacement dynamics.

## 5. Conclusions

Researching the origin and essence of bifurcation and chaos in urbanization process offers a new way of looking at complicated dynamics of simple systems. The pattern of phase space cannot be revealed by the 1-dimensional map based on ecological phenomena, but it can be brought to light by the 2-dimensional map based on rural-urban interaction. In this sense, cities and urbanization are good windows for us to investigate chaos and complexity. What is more, the analogy between urban systems and ecosystems will help us probe deeply into the essence of natural laws. By the study of urbanization dynamics, we can get three aspects of new knowledge about bifurcation and chaos. *First, the period-doubling bifurcation and chaos of determinate systems result from nonlinear interaction.* Because of the complicated behavior of the simple models such as logistic equation, bifurcation and chaos used to be thought of as inherent random behaviors of some determinate nonlinear models. However, the logistic model and the like usually suggest two-population interaction or even multi-population interaction. So it is anticipated that period-doubling bifurcation and chaos possibly root in nonlinear interaction, and the conclusion drawn from urban study can be generalized to ecological field. *Second, the chaotic behaviors of the logistic model do not indicate a chaotic attractor, and the relationship between chaos and fractals is scaling.* It is generally believed that chaos of the logistic process implies strange attractor with fractal structure before. However, the chaotic phase portrait of rural-urban population migration based on the interaction model suggests that the whole trajectory expands infinitely in the region between two radicals. Moreover, the chaotic spatial pattern follows the logarithmic function instead of the power law. The random distribution of the urban and rural data points in phase space shows no fractal structure. There seems to be no certain relation between chaos and strange attractor. However, both fractal structure and the route from bifurcation to chaos can be characterized by scaling process. *Third, the predator-prey interaction model can be revised to explain certain ecological phenomena, including logistic growth and oscillations.* A conjecture is that the essence of the complicated behaviors of the logistic equation in ecology is associated with the predator-prey interaction. Therefore, the classical Lotka-Volterra model can be reconstructed to interpret logistic process and periodic change (or fluctuation). Revising the coupling terms in the dynamical equations by normalization, the Lotka-Volterra model can be transformed into a new form. From the new predator-prey model, we can derive the logistic curve, oscillations and chaotic behavior. Where urban geography is concerned, the models of urbanization dynamics can be generalized to describe the spatial dynamics of urban morphology by means of fractal dimension growth. Moreover, both the models of urbanization and urban form evolution can be applied to developing the theory of spatial dynamics of replacement.

**Acknowledgements:** This research was sponsored by the National Natural Science Foundations of China (Grant No. 41590843 & 41671167). The support is gratefully acknowledged.